\documentclass[twocolumn,aps,pre,showpacs,amsmath,amsfonts,amssymb,floatfix,pre]{revtex4}
\usepackage{graphicx}
\usepackage{dcolumn}
\usepackage{bm}

\begin{document}

\title{How much random a random network is : a random matrix analysis}

\author{Sarika Jalan}
\email{sarika@mpipks-dresden.mpg.de}
\author{Jayendra N. Bandyopadhyay}
\email{jayendra@mpipks-dresden.mpg.de}
\affiliation{Max-Planck Institute for the Physics of Complex Systems, 
N\"{o}thnitzerstr. 38, D-01187 Dresden, Germany}

\begin{abstract}
We analyze complex networks under random matrix theory framework. Particularly,
we show that $\Delta_3$ statistic, which gives information about the long range 
correlations among eigenvalues, provides a qualitative measure of randomness in 
networks. As networks deviate from the regular structure, $\Delta_3$ follows random
matrix prediction of linear behavior, in semi-logarithmic scale with the slope of 
$1/\pi^2$, for the longer scale.
\end{abstract}

\maketitle

Triggered by recently available data on large real world networks (e.g. structure of the 
Internet or molecular networks in the living cell) combined with increasing computer 
power, an avalanche of qualitative research on network structure and dynamics are 
currently stimulating diverse scientific fields such as food web, nervous systems, 
cellular metabolism, scientific collaborations, Internet, human language etc. 
\cite{Strogatz,Barabasi,Boccaletti}. Results obtained so far span areas from the 
prevention of computer viruses to the stability and diversity of systems such as 
Internet, regulatory circuits of genome and ecosystems. These real world networks have 
several universal features, like small diameter, large clustering coefficient, scale-free 
degree distribution, assortative or disassortative mixing of the nodes, module structures 
\cite{module}, etc. Irrespective of real world networks having one or more above 
mentioned features, one thing is common in all of them, and that is the existence of some 
amount of {\it randomness or disorder} in the connections structure. According to many 
recent studies, randomness in connections is one of the most important and desirable 
ingredient for the proper functionality or the efficient performance of systems having 
underlying network structure \cite{Strogatz,face,language,game,Costa1}. For instance, 
information processing in brain is considered to be highly influenced by random 
connections among different modular structure \cite{face}, other examples of recent 
studies suggest the importance of randomness in various fields, such as evolution of 
language capacities of different species \cite{language}, evolution of cooperation in 
game theory \cite{game}, etc. On one hand these studies emphasize on the randomness in 
systems, on the other hand importance of {\it structure or regularity} is known for 
functional performance \cite{module}.

Seminal work of Watts and Strogatz shows that a very small amount of randomness has 
drastic impact on the diameter of network, leading to the so called small-world (SW) 
phenomenon \cite{SW}. Starting with a regular one dimensional lattice, random rewiring of 
connections leads to the SW behavior for very small rewiring probability $p$. This simple 
rewiring scheme preserves the regular lattice structure yielding high clustering 
coefficients, while increases in the number of random connections among distant nodes 
resulting in the decrease of diameter of the network. Though this model network captures 
two most important characteristics of the real world networks, the whole structure of the 
network still remains a very regular {\it one}-d lattice type, whereas real world 
networks are far more complex with a large number of random connections or {\it 
seemingly} random connections. The question arises whether one can identify or characterize 
the level of randomness in the complex networks ? There could be various possible ways to 
look into the problem. Recently L. da F. Costa has attempted to characterize randomness by 
looking for regular patterns in networks \cite{Costa2}. The purpose of present work is to 
have a qualitative measure of randomness in networks using eigenvalue fluctuation statistics of underlying adjacency matrix.

Eigenvalues distributions have been studied profusely in the literature 
\cite{book-Adj,Spectra}. They have certain signatures of the underlying network 
structures, like for complete random network it follows Wigner semi-circular law, for 
scale-free it follows triangular shape and for the small-world networks spectral 
distribution has multi-peaks structure \cite{Spectra,pap1}. In this paper we are 
interested in comparing the randomness in the network structure and except for the above 
mentioned typical network structures density distributions do not contain much 
informations about the network structures, particularly it is not able to address the 
issue of ``{\it how much randomness}".
 
Our earlier contributions \cite{pap1,pap2,pap3} have shown that the spacing distributions 
of complex random networks follow universal behavior of random matrix theory (RMT). In 
this paper we characterize the level of randomness in networks by using spectral rigidity 
test of RMT. Range for which $\Delta_3$ statistics follows random matrix prediction 
increases with the increase in random connections in the network.

RMT was proposed by Wigner to explain the statistical properties of nuclear spectra 
\cite{mehta}. Later this theory was successfully applied in the study of different 
complex systems including disordered systems, quantum chaotic systems, spectra of large 
complex atoms, etc \cite{rev-rmt}. More recently, RMT has been applied successfully to 
analyze time-series data of stock-market \cite{rmt-stock1,rmt-stock2}, atmosphere 
\cite{rmt-atmosphere}, human EEG \cite{rmt-brain}, and many more \cite{rev-rmt,Qgraph}. 
Nearest neighbor spacing distribution (NNSD) of the eigenvalues follows two universal 
properties depending upon the underlying correlations among the eigenvalues. For 
correlated eigenvalues, the NNSD follows Wigner-Dyson formula of Gaussian orthogonal 
ensemble (GOE) statistics of RMT; whereas, the NNSD follows Poisson statistics of RMT for 
uncorrelated eigenvalues.

We denote the eigenvalues of the adjacency matrix of network by 
$\lambda_i,\,\,i=1,\dots,N$, where $N$ is the size of the network and $\lambda_{i+1} > 
\lambda_i \forall \, \, i$. Eigenvalues are unfolded using the technique described in 
\cite{rev-rmt,pap1,pap2}. Using the unfolded spectra $\{\overline{\lambda}_i\}$, we 
calculate the nearest-neighbor spacings as
\begin{equation} 
s^{(i)}=\overline{\lambda}_{i+1}-\overline{\lambda}_i; \nonumber 
\end{equation} 
The NNSD $P(s)$ for the case of GOE statistics 
\begin{equation} 
P(s)=\frac{\pi}{2} s\exp \left(-\frac{\pi s^2}{4}\right). 
\label{nnsd}
\end{equation} 
The NNSD reflects only local correlations among the eigenvalues. The spectral rigidity, 
measured by the $\Delta_3$-statistic of RMT, provides information about the long-range 
correlations among the eigenvalues and is more sensitive test for RMT properties of the 
matrix under investigation \cite{mehta,casati}. In the following, we describe the 
procedure to calculate this quantity.

The $\Delta_3$-statistic measures the least-square deviation of the spectral staircase 
function representing the cumulative density $N(\overline{\lambda})$ from the best 
straight line fitting for a finite interval $L$ of the spectrum, i.e.,
\begin{equation}
\Delta_3(L; x) = \frac{1}{L} \min_{a,b} \int_x^{x+L} \,\left[
N(\overline{\lambda}) - a \overline{\lambda} -b \right]^2\,d \overline{\lambda}
\label{delta3}
\end{equation}
where $a$ and $b$ are obtained from a least-square fit. Average over several choices of $x$
gives the spectral rigidity $\Delta_3(L)$. For GOE, $\Delta_3(L)$ depends {\it logarithmically} 
on $L$, i.e.,
\begin{equation}
\Delta_3(L) \simeq \frac{1}{\pi^2} \ln L.
\label{del3_GOE}
\end{equation}

\begin{figure}
\centering
\includegraphics[width=9cm,height=10cm]{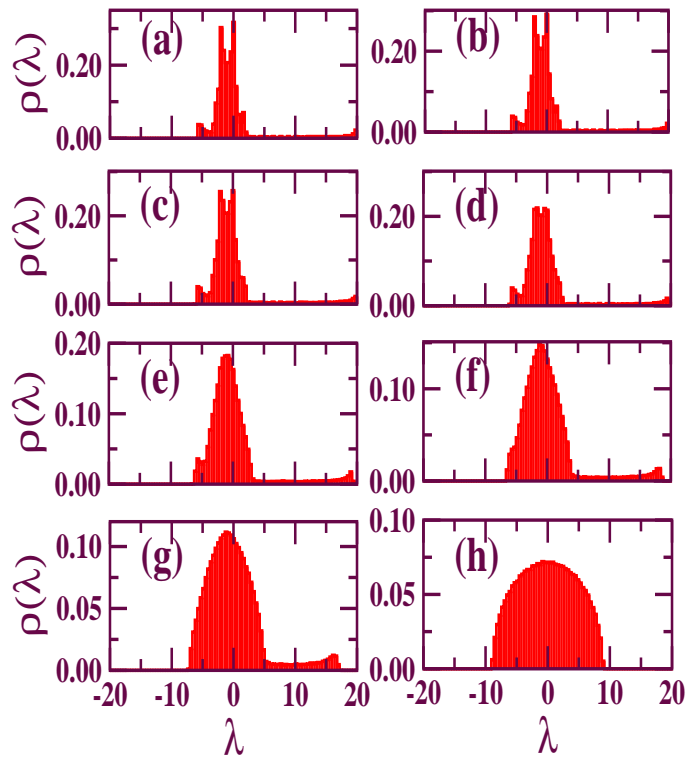}
\caption{(Color online) Figure shows changes in spectral density with the increase in random connections 
in the network. Initially, for (a)-(d) $p = 0.002, \,\,0.005,\,\,0.01$ and $0.02$ : 
spectral density show typical small-world network like distribution without having any 
specific analytical form with multiple peaks. (e)-(g) $p = 0.05, \,\,0.1$ and $0.2$ : 
approaching towards well known Wigner-Dyson semicircular distribution of random network, 
but having very long tail for larger positive eigenvalues. (h) $p = 1.0$ : completely 
random network, showing exact Wigner-Dyson semicircular distribution. All networks have 
$N=2000$ nodes and an average degree $k=20$ per node. Figures are plotted for the average 
over 10 random realizations of networks. }
\label{fig_density}
\end{figure}

Following we describe the method to generate networks with different randomness. Starting 
with one dimension ring lattice of $N$ nodes, in which every node is connected to its 
$k/2$ nearest neighbors, we randomly rewire each connection of the lattice with the 
probability $p$ such that self and multiple connections are excluded. Thus $p=0$ 
corresponds to a regular network, and $p=1$ gives a completely random network. The 
typical small-world behavior is observed around $p_c=0.002$ \cite{pap1}. Our earlier 
papers \cite{pap1,pap3} show that at this value of $p=p_c$ NNSD  
follows universal GOE prediction of RMT. Universal 
GOE behavior suggests that there exists a {\it minimal} amount of randomness in the 
networks yielding to the short range correlations among the corresponding eigenvalues. 
With the increase in $p$, obviously randomness increases in the network, but NNSD is not able 
to provide this information of enhancement of randomness.

As discussed in the introduction, NNSD only gives the information about the correlations 
among the neighboring eigenvalues. It does not tell anything about the long range 
correlations among the eigenvalues. We probe long range correlations among the 
eigenvalues using $\Delta_3$ statistic, which tells that how closely the network follows 
{\it ideal} behavior of GOE of random matrices.
For different values of $p \ge p_c$, we construct several networks of size $N=2000$ and 
average degree $k=20$ With increase in the value of $p$, 
number of random rewired connections increases. We study the spectral rigidity of 
networks generated for various $p$ and then present the results for ensemble average of 
networks for each $p$ value.

\begin{figure*}
\includegraphics[width=2\columnwidth,height=10cm]{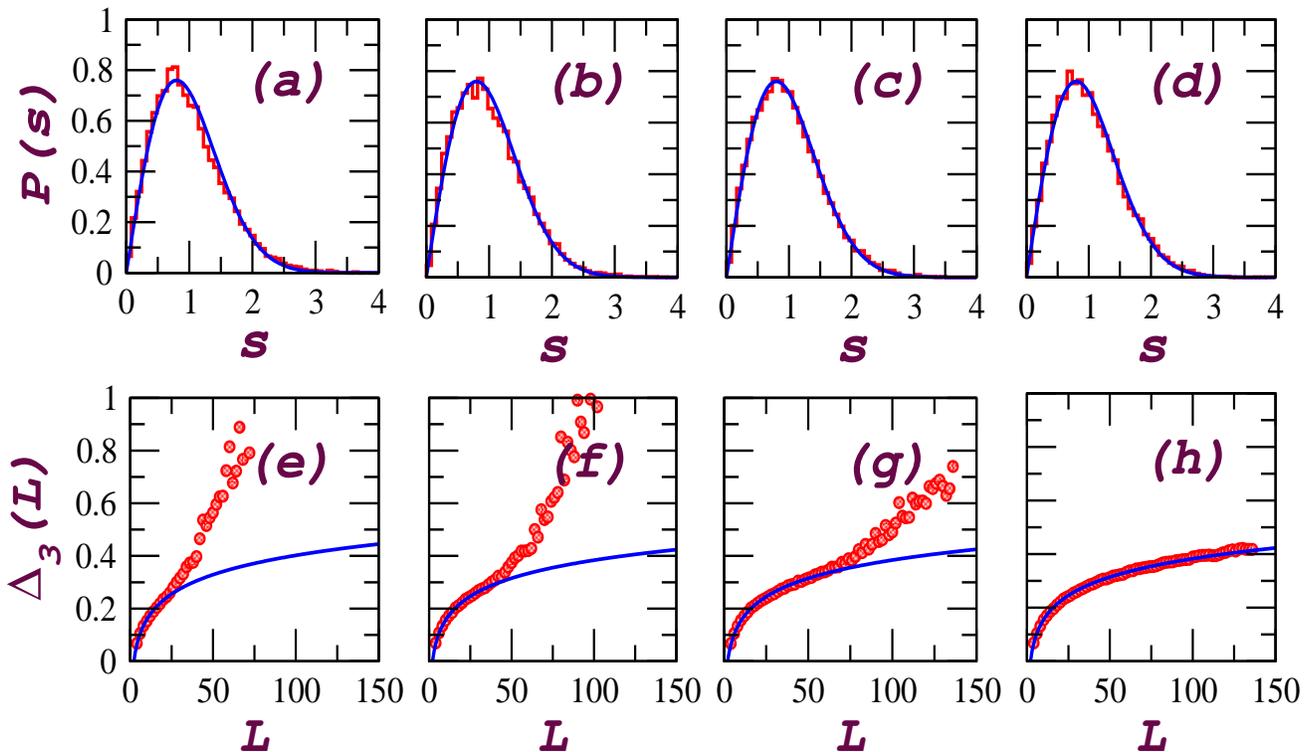}
\caption{(Color online) Figure shows changes in spectral behavior for different values of $p$. 
Network parameters remain same as for (\ref{fig_density}). (a) $p = 0.002$ : on the verge of the 
small-world transition, NNSD follows RMT predictions, $\Delta_3(L)$ statistic following 
RMT prediction only upto $L \sim 30$. (b) $p=0.01$ : spectral density 
(\ref{fig_density}(b)) and NNSD show similar behavior as for the previous $p$ value while 
$\Delta_3(L)$ statistic following RMT prediction for larger scale, upto $L \sim 40$. (c) 
$p=0.02$ :  $\Delta_3(L)$ statistic following RMT prediction upto $L \sim 75$. (d)$p = 
0.05$ : for such a small value of $p$, $\Delta_3(L)$ statistic following RMT prediction 
for very large scale $(L \sim 135)$. For $N=2000$, we can not have meaningful $\Delta_3$ 
statistic for the larger $L$ (see text). Hence for $p > 0.05$, $p = 0.1 \,\,0.2 
\,\,\mbox{and}\,\, 1.0$: $\Delta_3(L)$ statistic following RMT prediction for the scale 
$L \sim 135$.}
\label{fig_delta3} 
\end{figure*} 

We start with the network generated for $p = p_c = 0.002$, below 
this value of $p$ spectral properties of networks can not be modeled by GOE statistics of 
RMT (see our earlier papers \cite{pap1,pap2}). At $p=p_c$, there exists {\it minimal} 
amount of randomness sufficient to create correlations among the eigenvalues yielding GOE 
statistics of NNSD. Figure~\ref{fig_density} plots spectral distribution of several 
networks generated by Watts-Strogatz algorithm with different $p \ge p_c$ values. As $p$ 
increases, local regular structure destroys and random connections among nodes increase.

Figure~\ref{fig_delta3} plots NNSD distribution and $\Delta_3$ statistic of 
eigenvalues spectra of these networks. $\Delta_3(L)$ is calculated following 
Eq.~(\ref{delta3}) respectively. It can be seen from the plots in Figs. 
\ref{fig_delta3}(a)-(d), that after $p = p_c$ NNSD remains same with universal GOE 
statistics (\ref{nnsd}), and we do not infer anything more about the 
randomness in the network.  Though NNSD does not address the question of {\it how much 
randomness}, it contains a very important information; that various other aspects of RMT 
can now be applied to study these networks.

Figures~\ref{fig_delta3} (e)-(h) plot $\Delta_3$ statistic of several networks with 
different $p$ values. As it is seen from the figures $\Delta_3$ statistic has consistent 
different behavior for different $p$ values. Figs ~\ref{fig_delta3}(a)-(d) are 
plotted in the increasing order of the $p$ values, with (a) being plotted at the 
transition to the small-world behavior.  It can be seen from Figures~\ref{fig_density} 
(a) - (c), that the density distribution of eigenvalues for $p$ values $(0.002 - 0.02)$ 
are very similar with typical multi-peaks type of structure shown by small world 
networks. One can not infer from these figures that which network has high random 
connections and which one has less. NNSD plots are also same for these networks. But 
interesting information comes out when we look at the spectral rigidity plots.

Figure~\ref{fig_delta3}(e) shows that the spectral rigidity follows GOE predictions upto 
range $L \sim 30$. As $p$ is increased, the scale for which spectral rigidity follows GOE 
prediction, in general, increases. Figure~\ref{fig_delta3}(f) is plotted for $p=0.01$, 
for this value of $p$ spectral distribution and NNSD remains same for such a small 
difference of randomness from the previous $p$ values, but spectral rigidity now shows a 
bit different length scale ($L \sim 45$) for which it follows GOE predictions. Note that 
all the figures are plotted for the ensemble average of 10 networks generated for each 
$p$ values, so the statistics is the generic behavior of the networks with that 
randomness. For $p$ taking value $0.02$, Figure~\ref{fig_delta3} (g) shows that there 
exists a considerable change in the length scale for which spectral rigidity follows GOE 
prediction upto $L \sim 75$. There is a consistent increase in the value of $L$ with the 
$p$, indicating that $\Delta_3$ provides good measure of randomness in the networks. 
Figure~\ref{fig_delta3}(h) is plotted for $p = 0.05$, at this $p$ network follows GOE 
prediction till very long range $L \sim 135$. It means that eigenvalues which are $L=135$ 
distance apart, are also correlated. Note that as one probes for the larger range 
correlations, the statistical error in the calculation of $\Delta_3$ increases 
\cite{DysonIV}, and for the size of network ($N=2000$) we are considering here it becomes 
difficult to provide a meaningful test of RMT using this statistics. For $p > 0.05$, 
though spectral densities change (Figure~\ref{fig_density}(f)-(h)) being semicircular at 
$p \sim 1$, the spectral rigidity plots (Figs.\ref{fig_delta3}) (f)-(h) remains same as for $p 
\sim 0.05$. According to the RMT interpretation, eigenvalues of the network for $p \sim 
0.05$ are as much correlated as the network for $p \sim 1$, and for $p = 0.05$ network 
has as much {\it randomness} as in a complete random network.

We study complex networks under random matrix theory framework and use $\Delta_3$ 
statistics to provide a measure of deviation from the regular structure. Starting with a 
regular lattice, i.e. a network with complete regular structure, some connections are 
randomly rewired with probability $p$. Larger values of $p$ indicate more deviation from 
regular structure in the network. In this manner several networks are generated with 
different deviations from the regular structure. According to the RMT : $\Delta_3$ 
statistic, which measures long range correlations among the eigenvalues, provides 
insight about the randomness in the corresponding matrix. For a complete random matrix, 
eigenvalues are correlated till very long range. We use this result of eigenvalues 
correlation statistics to distinguish the randomness among the networks, and show that 
$\Delta_3$ can be used as a qualitative measure of randomness. With the increase in the 
value of $p$, the length scale for which $\Delta_3$ follows GOE prediction of RMT also 
increases. In the semi-logarithmic plots, the slope of $\Delta_3$ (Eq.~\ref{del3_GOE}) 
matches exactly the GOE predictions of $\sim 1/\pi^2$.

Interestingly, for the networks generated with $p \sim 0.05$, $\Delta_3$ statistic 
already starts following GOE prediction for the scale as long as for the networks with $p 
\sim 1$. According to the RMT, this tells that the eigenvalues of the networks generated 
with $p \sim 0.05$ are as much correlated as for the networks generated with $p \sim 1$, 
which are complete random networks. We interprete this result as following: for $p \sim 
0.05$ there is some kind of {\it spreading of randomness over the network} yielding the 
correlations among the eigenvalues as large as for the networks with $p=1$.

To conclude, continuing our RMT analysis of complex networks, this paper uses $\Delta_3$ 
statistic to study the randomness in the networks. It is shown that RMT is particularly 
suitable to study this important aspects of complex networks, thus on one hand enhancing 
the applicability of RMT, on the other hand opening a completely new domain to the complex 
networks studies. We hope that the complexity of the systems \cite{May} having network 
structures can be addressed under the RMT framework, and thus helping to understand the 
dynamical behavior \cite{book-syn} and robustness of such networks better \cite{SJ}.

SJ acknowledges T. Kruger (Technische Universit\"at Berlin) for encouraging 
discussions on the importance of results from graph theory prospective, and 
M. M\"uller (currently at MPIPKS, Dresden) from random matrices prospective.

\end{document}